# When *Follow* is Just One Click Away: Understanding Twitter *Follow* Behavior in the 2016 U.S. Presidential Election

**Yu Wang**
Computer Science
University of Rochester
Rochester, NY 14627
ywang176@ur.rochester.edu

**Xiyang Zhang**
Psychology
Beijing Normal University
Beijing 100875
zxy2013@mail.bnu.edu.cn

**Jiebo Luo**
Computer Science
University of Rochester
Rochester, NY 14627
jluo@cs.rochester.edu

## Abstract

Motivated by the two paradoxical facts that the marginal cost of following one extra candidate is close to zero and that the majority of Twitter users choose to follow only one or two candidates, we study the Twitter *follow* behaviors observed in the 2016 U.S. presidential election. Specifically, we complete the following tasks: (1) analyze Twitter *follow* patterns of the presidential election on Twitter, (2) use negative binomial regression to study the effects of gender and occupation on the number of candidates that one follows, and (3) use multinomial logistic regression to investigate the effects of gender, occupation and celebrities on the choice of candidates to follow.

## Introduction

President Obama is often credited as the first to extend his political campaign onto social media during his first presidential run in 2008 (Tumasjan et al. 2010). Eight years later in the 2016 presidential election, social media was considered to be Donald Trump's most powerful weapon (Alaimo 2016; Lockhart 2016). After winning the election, Donald Trump himself commented that tweeting "is a great way of communication" (Stahl 2016). In his book *Our Revolution*, which reflects on the 2016 presidential campaign, Bernie Sanders suggests that one of the reasons why his campaign did well is the campaign team's success with social media (Sanders 2016). One opinion shared by both Trump and Sanders is that having a large number of followers on Twitter is an invaluable campaign asset (Stahl 2016; Sanders 2016).

Given the prominent role that Twitter played in the presidential election, a systematic study of how individuals behave on Twitter and the informing factors underlying the observed behavior is warranted. Our work is motivated (1) by the paradoxical observation that most individuals choose to follow only one or two presidential candidates, when the marginal cost of following the fifteen others is just one click away and therefore technically close to zero, and (2) by the common criticism that Twitter *follow* is not a strong signal of support. We started by compiling the entire universe of the 2016 U.S. presidential election on Twitter.

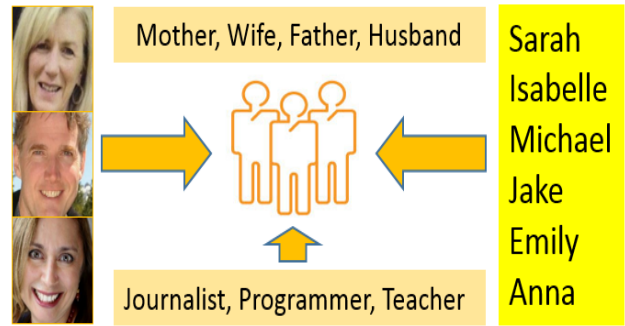

Figure 1: We use first names, profile images and family roles to identify gender, and we extract from self-descriptions individuals' occupations.

We recorded all the 15.5 million individuals who were following one or more of the 16 candidates in early April, 2016. With such a rich dataset, we are able to (1) explore the most frequent *follow* patterns among these 15.5 million individuals, (2) explore the correlation between following two different candidates and examine the question of electorate polarity, (3) study the effects of gender and occupation on the number of candidates that one chooses to follow with negative binomial regression, and (4) study the effects of gender, occupation and the endorsement of celebrities on the choice of candidates. When coding the variable *gender*, we integrate information from first names, profile images and self descriptions (Figure 1) and we show in the paper that the effect of gender is consistent across all the three channels.

The contributions of our paper are as follows: (1) we demonstrate to what extent the leading presidential candidates have dominated the Twitter sphere using a weighted follower metric, (2) we quantitatively measure how polarized Twitter followers are when it comes to choosing presidential candidates across party lines, (3) we show that women tend to follow fewer candidates than men and that journalists are more likely to follow a large number of candidates, (4) we find that women tend to follow Democratic candidates, which supports the idea that women vote following party lines (Miller 2016), and (5) we find that people who follow celebrities such Beyoncé, Lebron James and

Lady Gaga are more likely to be following candidates of both parties and that when choosing between Hillary Clinton and Bernie Sanders, endorsements by these celebrities favor Clinton.

## Related Literature

Our paper builds on previous literature on electoral studies, data mining and computer vision.

Previous work has studied the increasing polarization of American politics at both the elite level (Hare and Poole 2014; McCarty, Poole, and Rosenthal 2009) and the mass level (Campbell 2016; Doherty 2014). Druckman *et al.*, in particular, study how elite partisan polarization affects public opinion formation and find that party polarization decreases the impact of substantive information (Druckman, Peterson, and Slothuus 2013). Social clustering, on the other hand, is analyzed in (McPherson, Smith-Lovin, and Cook 2001; Barberá 2015). In our work, we contribute to analyzing political polarization at the public level on Twitter.

Gender plays an important role in the forming and dissolving of relationships (Burt 2000), in online behavior (Ottoni et al. 2013) and in political voting (King and Matland 2003; Dolan 2008; Brians 2005; Wang et al. 2016). One common observation is that women tend to vote for women, which is usually referred to as gender affinity effect. In this paper we will analyze specifically the effects of gender on the number of presidential candidates that an individual chooses to follow and on which party that one chooses to follow.

Given the importance of gender in real applications, a large number of studies have attempted to classify gender based on user names (Mislove et al. 2011; Nilizadeh et al. 2016), tweets, screen name and description (Burger et al. 2011) and friends (Zamal, Liu, and Ruths 2012). Following this line of research, our study will take advantage of information from both user names and user-provided descriptions.

Recent advances in computer vision (Krizhevsky, Sutskever, and Hinton 2012; Simonyan and Zisserman 2015; Srivastava, Greff, and Schmidhuber 2015; He et al. 2016), on the other hand, have made object detection and classification increasingly accurate. In particular, face detection and gender classification (Farfade, Saberian, and Li 2015; Jia and Cristianini 2015; Levi and Hassner 2015) have both achieved very high accuracy, largely thanks to the adoption of deep learning (LeCun, Bengio, and Hinton 2015) and the availability of large datasets (Huang et al. 2007; Jr. and Tesafaye 2006; Phillips et al. 1998) and more recently (Guo et al. 2016). Our paper extracts gender-related information based on Twitter profile images and is related to gender classification using facial features (Levi and Hassner 2015; Ginosar et al. 2015; Wang, Li, and Luo 2016; Wang et al. 2016; Nilizadeh et al. 2016).

## Data

Our dataset includes two components, both of which come from Twitter. The first component consists of the followers' Twitter ID information for all the presidential candidates in

Table 1: The Number of Followers (April, 2016)

| Candidate | # Followers | Candidate | # Followers |
|---|---:|---|---:|
| Chafee (D) | 23,282 | Clinton (D) | 5,855,286 |
| O'Malley (D) | 130,119 | Sanders (D) | 1,859,856 |
| Webb (D) | 25,731 | Bush | 25,731 |
| Carson (R) | 1,248,240 | Christie (R) | 120,934 |
| Cruz (R) | 1,012,955 | Fiorina (R) | 672,863 |
| Kasich (R) | 266,534 | Huckabee (R) | 460,693 |
| Paul (R) | 841,663 | Rubio (R) | 1,329,098 |
| Trump (R) | 7,386,778 | Walker (R) | 226,282 |

Note: Sorted by party affiliation and alphabetically.

April, 2016. This component is exhaustive in the sense that we have recorded all the followers' IDs. In total, there are 15,455,122 individuals following the 16 presidential candidates and some of them are following more than 1 candidate. We transform this component into a 15.5 million by 16 matrix of 1's and 0's, with each row representing an individual and each column a presidential candidate. We report the summary statistics in Table 1[1]. It can be easily observed that Donald Trump and Marco Rubio have the largest numbers of followers among the Republican candidates and that Hillary Clinton and Bernie Sanders have the largest numbers of followers among the Democratic candidates.

The second component of our dataset has 1 million individuals, randomly sampled from the first component.[2] Based on these individuals, we extract user name, user-provided description, the starting year of using Twitter, social capital (Wang et al. 2016), and the profile image (Wang, Li, and Luo 2016).

The third component comprises follower information of Beyoncé, Lady Gaga, Lebron James, three media celebrities all of whom have explicitly endorsed Hillary Clinton. These three celebrities constitute a significant presence among individuals who follow the presidentail candidates: 5.65% of the individuals in the dataset follow Beyoncé, 15.9% follow Lebron James and 19.58% follow Lady Gaga. In the book *Our Revolution*, Bernie Sanders also emphasizes the importance of celebrity support (Sanders 2016). Donald Trump, by contrast, contended that he does not need celebrities to fill up rallies, when Jay Z and Beyoncé held public events to rally votes for Clinton.[3]. This data component then enables us to analyze the celebrity effect in a quantitative manner: whether individuals who follow these celebrities are also more likely to follow the candidates who have won the celebrities' endorsement.

We summarize the variables used in this work and their definitions in Table 2.

---

[1] The number of followers is changing every minute. To make sure the statistics are comparable across candidates, we decided to collect the data exclusively between April 1st and April 7th, 2016.

[2] To faciliate replication of our results, we have set the random seed (Python) to 11.

[3] https://www.washingtonpost.com/video/politics/trump-we-dont-need-jay-z-to-fill-up-arenas/2016/11/05/25d536e2-a365-11e6-8864-6f892cad0865_video.html.

Table 2: Variable Definitions

| Name | Definition |
|---|---|
| Independent variables: | |
| Tweets | Count, number of tweets posted |
| Social Capital | Count, number of followers |
| Journalist | Binary, a journalist |
| Name | Binary, female based on first names |
| Image | Binary, female based on profile images |
| Description | Binary, female based on self-provided descriptions |
| Female | Binary, female by first name or image or description. |
| Beyoncé | Binary, follow Beyoncé |
| Lady Gaga | Binary, follow Lady Gaga |
| Lebron James | Binary, follow Lebron James |
| Celebrity | Binary, follow Beyoncé or Lady Gaga or Lebron James |
| Dependent variables: | |
| # Candidates | Count, number of candidates that one follows |
| Democrat follower | Binary, follow Democrats only |
| Republican follower | Binary, follow Republicans only |
| Independent follower | Binary, follow both Democrats and Republicans |
| Bernie Sanders | Binary, follow Bernie Sanders |
| Hillary Clinton | Binary, follow Hillary Clinton |

Note: (1) Following previous studies (Wang et al. 2016), we define social capital on Twitter as the raw number of followers. (2) By construction, Democrat follower, Republican follower and Independent follower always sum up to 1.

## Methodology

### Gender classification

We employ three methods to extract information on gender. As in several prior studies (Mislove et al. 2011; Nilizadeh et al. 2016), we first compile a list of 800 names, based on appearance frequency on Twitter, that are gender-revealing, such as Mike, Jake, Emily, Isabella and Sarah.[4] This constitutes our first channel. We then use this list to classify individuals whose names are contained in this list. As one would expect, a large number of individuals can not be classified with this list.

Our second channel is the profile image. We train a convolutional neural network using 42,554 weakly labeled images, with a gender ratio of 1:1. These images come from Trump's and Clinton's followers. We infer their labels using the followers' names (channel 1). For validation, we use a manually labeled data set of 1,965 profile images for gender classification. The validation images come from Twitter as well so we can avoid the cross-domain problem. Moreover, they do not intersect with the training samples as they come exclusively from individuals who unfollowed Hillary Clinton before March 2016.

[4]The complete name list is available for download on the first author's website.

The architecture of our convolutional neural network is reported in Figure 2, and we are able to achieve an accuracy of 90.18%, which is adequate for our task (Table 3).[5]

Table 3: Summary Statistics of CNN Performance (Gender)

| Architecture | Precision | Recall | F1 | Accuracy |
|---|---|---|---|---|
| 2CONV-1FC | 91.36 | 90.05 | 90.70 | 90.18 |

Third, we extract gender-revealing keywords from user-provided descriptions. These keywords are *papa, mama, mom, father, mother, wife* and *husband*.

We prioritize the first channel (first names) most and the third channel (self description) the least. Only when the more prioritized channels are missing do we use the less prioritized channels:

**first names > profile images > self descriptions**

Based on this ranking, we are able to label 38.7% of the observations from first names, another 17.2% with profile images and 0.7% with self descriptions. In total, we are able to classify 56.6% of the 1 million individuals. We summarize the number of labeled individuals and the net contribution of each channel in Table 4.

Table 4: 3-Channel Classification of Gender

| Channel | First Name | Profile Image | Self Description |
|---|---|---|---|
| Priority | 1 | 2 | 3 |
| Identification | 387,148 | 304,278 | 30,786 |
| Contribution | 38.7% | 17.2% | 0.7% |

Note: Partly as a result of our priority ranking, the net contribution of profile images is significantly smaller than first names. The net contribution of self descriptions (3rd channel) is about 1 percent.

### Negative binomial regression

Our work is motivated by the observation that the majority of individuals choose to follow only one or two candidates when the marginal cost of "following" other candidates is just one click away. In order to understand this phenomenon, especially the role that gender plays, we apply the negative binomial regression (Greene 2008) and link the number of candidates that one follows, which is count data, to the explanatory gender variable. In this regression, the conditional likelihood of the number of candidates that individual $j$ follows, $y_j$, is formulated as

$$f(y_j|v_j) = \frac{(v_j\mu_j)^{y_j} e^{-v_j\mu_j}}{\Gamma(y_j+1)}$$

where $\mu_j = exp(\mathbf{x_j}\boldsymbol{\beta})$ is the link function that connects our explanatory variables to the number of candidates that one chooses to follow and $v_i$ is a hidden variable with a Gamma($\frac{1}{\alpha}, \alpha$) distribution. After plugging in the explanatory variables, the unconditional log-likelihood function takes the form:

$$\ln L = \sum_{j=1}^{N}[ln(\Gamma(m+y_j)) - ln(\Gamma(y_j+1)) - ln(\Gamma(m))$$
$$+ m ln(p_j) + y_j ln(1-p_j)]$$
$$p = 1/(1+\alpha\mu)$$
$$m = 1/\alpha$$
$$\mu = exp(\beta_0 + \beta_1 \text{Tweets Posted} + \beta_2 \text{Follower Count}$$
$$+ \beta_3 \text{Journalist} + \beta_4 \text{Year Fixed Effects}$$
$$+ \boldsymbol{\beta_5} \cdot \textbf{Name} + \boldsymbol{\beta_6} \cdot \textbf{Image}$$
$$+ \boldsymbol{\beta_7} \cdot \textbf{Description} + \boldsymbol{\beta_8} \cdot \textbf{Female})$$

where $\alpha$ is the over-dispersion parameter and will be estimated as well.

[5]The trained model has been deployed at our demo website.

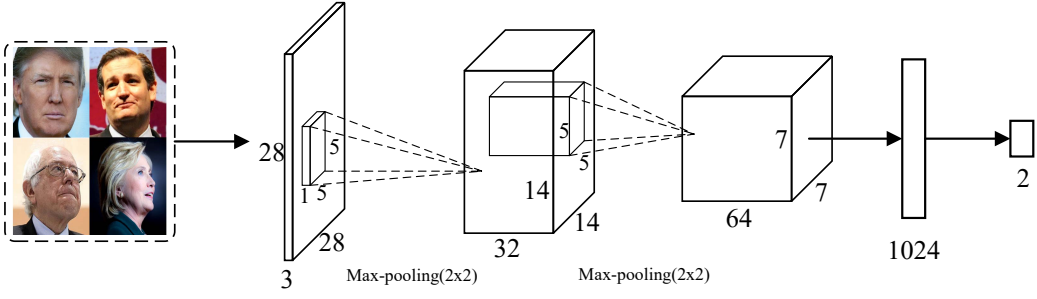

Figure 2: The CNN model consists of 2 convolutional layers, 2 max-pool layers, and a fully connected layer.

## Multinomial Logistic Regression

Besides the number of candidates, another question we try to answer is which candidates one chooses to follow. For this purpose, we identify three classes: (1) follow Democratic candidates only, (2) follow candidates from both parties, and (3) follow Republican candidates only. We use the class $c$ as the dependent variable and formulate the probability of each observation in a multinomial logistic setting (Maddala 1983):

$$P1 = Pr(c=1) = \frac{e^{\mathbf{x}\beta_1}}{e^{\mathbf{x}\beta_1} + 1 + e^{\mathbf{x}\beta_3}}$$
$$P2 = Pr(c=2) = \frac{1}{e^{\mathbf{x}\beta_1} + 1 + e^{\mathbf{x}\beta_3}}$$
$$P3 = Pr(c=3) = \frac{e^{\mathbf{x}\beta_3}}{e^{\mathbf{x}\beta_1} + 1 + e^{\mathbf{x}\beta_3}}$$

where $\mathbf{x}$ is the vector of explanatory variables: number of posted tweets, number of followers, being a journalist (binary), gender, following a celebrity and year controls. Notice that the coefficients for the second class (following candidates from both parties) have been normalized to 0 to solve the identification problem.

The log-likelihood function then takes the form:

$$\ln L = \sum_{i=1}^{n}[\delta_{1i}\ln(P1) + \delta_{2i}\ln(P2) + \delta_{3i}\ln(P3)]$$

where $\delta_{ij}$=1 if i=j and 0 otherwise. Note that logistic regression, which we will use to differentiate the celebrity effects on Hillary Clinton and Bernie Sanders, is a special case of the multinomial logistic regression with $\beta_3$ set to zero.

## Results

In this section, we report on (1) election *follow* patterns observed on Twitter (2) negative binomial regression analysis of the number of candidates that one follows (3) multinomial logistic regression analysis of gender affinity effects on the choice of candidates and (4) logistic regression analysis of celebrity effects.

### Election *Follow* Patterns on Twitter

In Table 5, we report on how committed each candidate's followers are. By commitment, we mean how many of the followers follow only that one specific candidate. It can be seen that Clinton, Trump and Sanders have highest percentages of 'committed' followers in the Twitter sphere, whereas only 9 percent of Bush's 529,820 followers follow him alone and 89 percent of Cruz's 1,012,955 followers follow other candidates besides Cruz. This suggests that while having a large number of followers is always beneficial, not all followers are equally committed.

To overcome this problem, we propose a simple and intuitive method to weight each follower by the reciprocal of the total number of candidates that he or she is following. For example, an individual who follows Bernie Sanders, Donald Trump and Ted Cruz will receive a weight of $\frac{1}{3}$, and an individual who follows Hillary Clinton only will receive a weight of 1. Mathematically, the Twitter share of candidate $j$ is then calculated as:

$$\text{share}_j = \frac{\sum_{i=1}^{n}\delta_{ij}\text{weight}_i}{\sum_{k=1}^{m}\sum_{i=1}^{n}\delta_{ik}\text{weight}_i}$$
$$\text{weight}_i = \frac{1}{\sum_{k=1}^{m}\delta_{ik}}$$

Table 5: Follower Engagement for Each Candidate (in Decimals)

| Candidate | # 1 | # 2 | # 3 | # 4 | # 5+ |
|---|---|---|---|---|---|
| Chafee | 0.39 | 0.13 | 0.08 | 0.06 | 0.34 |
| Clinton | 0.75 | 0.16 | 0.04 | 0.02 | 0.04 |
| O'Malley | 0.29 | 0.23 | 0.17 | 0.07 | 0.24 |
| Sanders | 0.6 | 0.23 | 0.07 | 0.03 | 0.07 |
| Webb | 0.15 | 0.13 | 0.1 | 0.09 | 0.52 |
| Bush | 0.09 | 0.16 | 0.14 | 0.11 | 0.51 |
| Carson | 0.24 | 0.19 | 0.14 | 0.12 | 0.3 |
| Christie | 0.11 | 0.13 | 0.12 | 0.09 | 0.56 |
| Cruz | 0.11 | 0.17 | 0.18 | 0.15 | 0.39 |
| Fiorina | 0.41 | 0.15 | 0.08 | 0.07 | 0.29 |
| Kasich | 0.25 | 0.13 | 0.1 | 0.08 | 0.44 |
| Huckabee | 0.28 | 0.15 | 0.11 | 0.09 | 0.37 |
| Paul | 0.2 | 0.16 | 0.14 | 0.12 | 0.38 |
| Rubio | 0.25 | 0.16 | 0.15 | 0.13 | 0.32 |
| Trump | 0.72 | 0.14 | 0.05 | 0.03 | 0.06 |
| Walker | 0.14 | 0.08 | 0.08 | 0.08 | 0.62 |

Note: '#5+' stands for 'following five or more presidential candidates. For example, six percent of Trump followers follow five or more candidates.

where n is the total number of followers (15,455,122), m is the total number of candidates (16), $\delta_{ik}$ is 1 if individual i follows candidate k and 0 otherwise.

After applying this weighting mechanism, we find the Twitter share of the leading candidates, such as Donald Trump, Hillary Clinton and Bernie Sanders, further increases. Their aggregated share of Twitter followers rises from 68.7% to 80.1% (Figure 3).

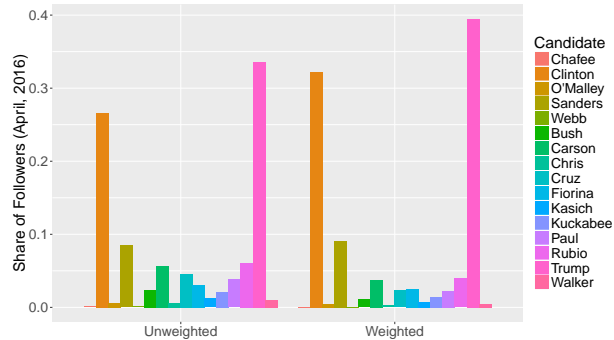

Figure 3: Share of the three leading candidates Trump, Clinton and Sanders further increases after weighting the followers.

We further analyze the top 15 most frequent patterns present in the Twitter sphere (Table 6). One immediate observation is that Trump, Clinton and Sanders are the three dominant forces in the Twitter sphere. 34.5% of the individuals recorded in our exhaustive dataset are following Donald Trump alone. 28.4% are following Hillary Clinton alone. 7.2% are following Sanders alone. These three groups account for 69.9% of the entire recorded population in our dataset. Individuals who follow only Marco Rubio or Carson or Fiorina make up no more than 2 percent of the population. Individuals who follow both Clinton and Trump constitute 3 percent of the entire recorded population.[6] Other frequent 2-itemsets (Han, Kamber, and Pei 2011) include Carson and Trump (1%), Sanders and Trump 0.6/% and Rubio and Trump (1%). The only 3-itemset among the top 15 frequent pattern is Clinton, Sanders and Trump (0.5%).

Table 6: Top 15 Most Frequent Items in the Election's Twitter Sphere

| 1 | 0.345 | Trump |
|---|---|---|
| 2 | 0.284 | Clinton |
| 3 | 0.072 | Sanders |
| 4 | 0.030 | Clinton Trump [2-itemset] |
| 5 | 0.021 | Rubio |
| 6 | 0.021 | Clinton Sanders [2-itemset] |
| 7 | 0.020 | Carson |
| 8 | 0.018 | Fiorina |
| 9 | 0.011 | Paul |
| 10 | 0.010 | Carson Trump [2-itemset] |
| 11 | 0.008 | Kuckabee |
| 12 | 0.007 | Cruz |
| 13 | 0.006 | Sanders Trump [2-itemset] |
| 14 | 0.006 | Rubio Trump [2-itemset] |
| 15 | 0.005 | Clinton Sanders Trump [3-itemset] |

We further examine how the decision of following one candidate correlates with that of following another candidate using the Pearson correlation coefficient. One immediate observation is that correlation between following candidates from the same party tends to be positive and correlation between following candidates from different parties tends to be negative (Figure 4). In particular, the correlation is -0.51 between Clinton and Trump and -0.22 between Sanders and Trump. By contrast, Marco Rubio and Ted Cruz have a strong and positive correlation coefficient of 0.43. This constitutes our first piece of evidence that individuals on Twitter are also polarized (Campbell 2016).

Motivated by the fact that Twitter *follow* behavior appears to cluster around the two parties, we refer to individuals who follow Democratic candidates exclusively as *Democrat followers* and refer to individuals who follow Republican candidates exclusively as *Republican followers* and lastly we refer to those who follow candidates from both parties as *Independent followers*.[7] It turns out that 92% of the 15.5 million followers are either Democrat followers or Republican followers, i.e., they are following candidates from only one party not both parties, which lends further support to the idea that the public are polarized on Twitter (Campbell 2016).

---

[6]This number is surprisingly low and suggests that Twitter 'follow' behavior is more of a signal of support/interest than communication as far as the presidential campaign is concerned.

[7]Note that this definition is based on Twitter *follow* behavior not on real party affiliation.

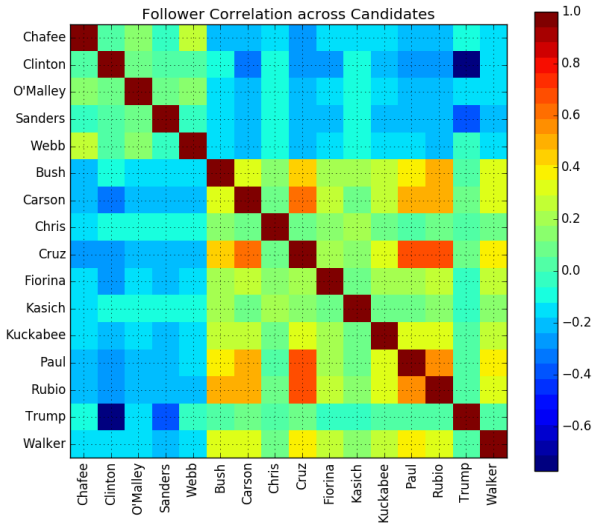

Figure 4: Party clustering observed in Twitter following behavior. Individuals who follow Trump are more likely to follow Ted Cruz and Marco Rubio and less likely to follow Hillary Clinton or Bernie Sanders.

### Negative Binomial: Follow the Candidates

Having summarized the election $follow$ patterns as a whole, we are now ready to analyze the factors behind an individual's decision to follow a certain number of candidates. In particular, while the marginal cost of following an extra candidate is close to zero, most individuals choose to follow only 1 or 2 candidates (Figure 5).

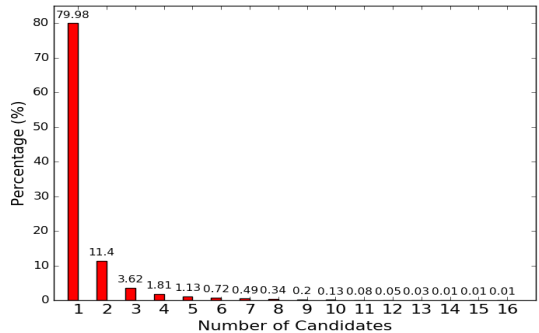

Figure 5: This figure is generated using all 15.5 million observations. In spite of the low marginal cost of following extra people on Twitter, most individuals chose to follow no more than 2 candidates during the 2016 U.S. presidential election.

In the regression, we use the number of candidates that one follows (# candidates) as the dependent variable. *social capital*, *journalist* and gender are the three variables that we are particularly interested in. The coefficient on *social capital* would enable us to learn whether more prominent individuals tend to follow more candidates or not.

The coefficient on *journalist* measures whether journalists tend to follow a larger number of the presidential candidates (and we expect the answer to be yes). *gender* measures the effects of being female. Following (Burt 2000; Wang et al. 2016), we expect the coefficient on *gender* to be negative, i.e., women tend to follow fewer candidates.

We report our regression results in Table 7. Across all the four specifications, we find that *tweets*, *social capital* and *journalist* are all positively correlated with the number of candidates that one chooses to follow. With respect to gender, we find that regardless of the channel that we use (name in Column 1, image in Column 2, description in Column 3, all the three in Column 4), the coefficient on *female* is consistently negative, suggesting that women are more likely to follow fewer candidates. [8]

Table 7: Negative Binomial: the Number of Candidates

|  | Name | Image | Descr. | All |
|---|---|---|---|---|
| # Candidates |  |  |  |  |
| Tweets | 2.906*** | 2.180*** | 3.543*** | 2.427*** |
|  | (0.146) | (0.134) | (0.418) | (0.105) |
| Social Capital | 2.168*** | 1.153*** | 0.753 | 1.282*** |
|  | (0.482) | (0.329) | (0.920) | (0.298) |
| Journalist | 0.249*** | 0.223*** | 0.0444 | 0.201*** |
|  | (0.0234) | (0.0208) | (0.0646) | (0.0181) |
| Name | -0.0817*** |  |  |  |
|  | (0.00270) |  |  |  |
| Image |  | -0.0286*** |  |  |
|  |  | (0.00340) |  |  |
| Description |  |  | -0.150*** |  |
|  |  |  | (0.00936) |  |
| Female |  |  |  | -0.0536*** |
|  |  |  |  | (0.00230) |
| Year F.E. | Yes | Yes | Yes | Yes |
| Constant | 0.257* | 0.268* | 0.546 | 0.299** |
|  | (0.129) | (0.134) | (0.319) | (0.108) |
| $\ln(\alpha)$ |  |  |  |  |
| Constant | -3.804*** | -4.327*** | -2.046*** | -4.320*** |
|  | (0.0441) | (0.0826) | (0.0330) | (0.0600) |
| Observations | 387148 | 294987 | 30786 | 557777 |

Standard errors in parentheses
* $p < 0.05$, ** $p < 0.01$, *** $p < 0.001$

### Multinomial Logistic Regression: Gender Affinity Effect

Having demonstrated that women behave differently from men in the number of candidates that they choose to follow, in this subsection we analyze whether women also differ from men in choosing *which* candidates to follow. We summarize the gender ratio of each candidates' followers in Figure 6.[9] It can be seen that Clinton has the highest female to male ratio, followed closely by Bernie Sanders. Rand Paul

---

[8] In all the specifications, we have controlled for the year fixed effects.

[9] We used all the three channels to extract gender-related information.

(R) and Jim Webb (D) on the other hand have the lowest female to male ratio. In general, the Democratic candidates mostly have a gender ratio close to or over 40%, while the Republican candidates tend to have a gender ratio well below 40%. Carly Fiorina, the only female candidate in the Republican party, is the only Republican to reach 40%.

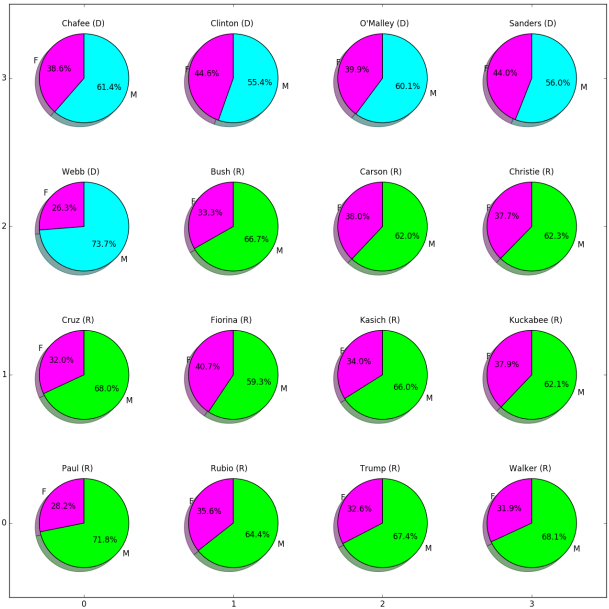

Figure 6: In percentage, the leading Democratic candidates have more female followers than the leading Republican candidates.

Building from previous studies (King and Matland 2003; Dolan 2008; Brians 2005; Wang et al. 2016), we construct a multinomial logistic regression model to test whether women are more likely to follow Democratic candidates. In addition, we examine whether followers of Beyoncé, Lady Gaga and Lebron James, all of whom have explicitly endorsed Hillary Clinton, thus revealing support for Democratic causes, are more likely to follow Democratic candidates exclusively.

We report our results in Table 7. Using *Independent* as the baseline for comparison, we examine the role of social capital, occupation, gender and celebrities. Across all specifications, we find that people with higher social capital and people working as a journalist are more likely to be *Independent* followers, i.e., following candidates from both parties.

From Columns 1 to 4, we examine the role of gender in determining whom to follow. The coefficient on gender is negative for Republican, positive for Democrat and 0 for Independent, suggesting that women are more likely to follow Democrats and less likely to follow Republicans (Figure 7). This result is consistent across all the four specifications.

From Columns 5 to 8, we examine the role of celebrities in determining whom to follow. The coefficient on celebrity (Beyoncé, Lebron James and Lady Gaga) is negative for both *Democratic follower* and *Republican follower* and 0 for *Independent follower*, suggesting that individuals who are following these celebrities are more likely to follow candidates from both parties. This result is also consistent across all the four specifications (5-8).

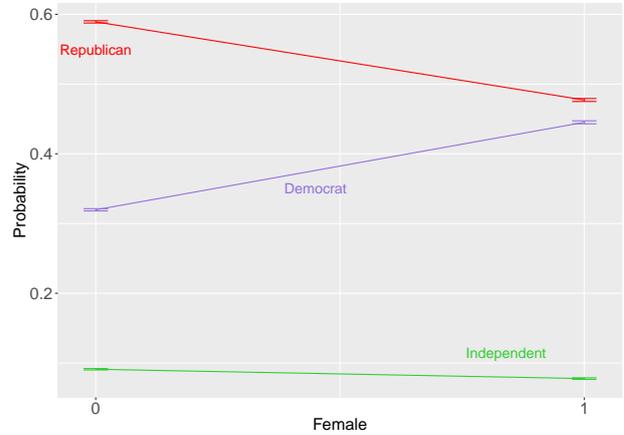

Figure 7: As indicated by the slope, the effects of being female are positive on Democrat, negative on Republican and slightly negative on Independent.

### Logistic Regression: Celebrity Effects

When comparing between *Democratic followers, Republican followers and Independent followers*, we find that individuals who follow celebrities are more likely to be *Independent followers*, i.e., they follow candidates from both parties. In this subsection, we restrict our comparison to the two dominant classes, Democratic followers and Republican followers, which constitute 92% of our observations. In particular, we will analyze whether following celebrities who have explicitly endorsed Hillary Clinton can affect (1) the probability of being a Democratic follower over a Republican follower and (2) the probability of following Hillary Clinton instead of Bernie Sanders.

We report our results in Table 8. When comparing between Democrat followers and Republican followers (Columns 1-5), we find that individuals who follow Beyoncé and Lady Gaga are more likely to be Democratic followers. By contrast, those who follow Lebron James are more likely to be Republican followers, which suggests that celebrities and celebrity followers do not necessarily share the same political opinion. In addition, we show in Column 3 that the interaction variable *James # Female* has a positive and significant coefficient, suggesting that effect of following Lebron James is significantly smaller for females than for males with regards to following the presidential candidates.

When restricting our observation to only Sanders followers and Clinton followers, we find that all the celebrity effects to be positive, suggesting that endorsements by these celebrities are giving Hillary Clinton an edge over Sanders for individuals who follow these celebrities.

Table 8: Multinomial Logistic Analysis of the Choice of Candidates to Follow

| | Names Only | Images Only | Description Only | All | Beyoncé | Lebron James | Lady Gaga | Celebrity |
|---|---|---|---|---|---|---|---|---|
| **Democratic Follower** | | | | | | | | |
| Tweets | -0.909 | -0.856 | 1.561 | -0.384 | -0.149 | -0.296 | -0.333 | -0.180 |
| | (0.605) | (0.543) | (1.707) | (0.446) | (0.452) | (0.450) | (0.447) | (0.451) |
| Social Capital | -5.104* | -6.716* | -1.351 | -7.334** | -6.521** | -7.145** | -6.831** | -6.757** |
| | (2.568) | (3.038) | (2.537) | (2.579) | (2.515) | (2.593) | (2.543) | (2.525) |
| Journalist | -0.749*** | -0.677*** | -0.536** | -0.640*** | -0.646*** | -0.668*** | -0.651*** | -0.664*** |
| | (0.0801) | (0.0686) | (0.206) | (0.0608) | (0.0609) | (0.0610) | (0.0609) | (0.0610) |
| Name | 0.644*** | | | | | | | |
| | (0.0128) | | | | | | | |
| Image | | 0.316*** | | | | | | |
| | | (0.0149) | | | | | | |
| Description | | | 0.732*** | | | | | |
| | | | (0.0426) | | | | | |
| Female | | | | 0.490*** | 0.516*** | 0.438*** | 0.506*** | 0.490*** |
| | | | | (0.0106) | (0.0107) | (0.0107) | (0.0106) | (0.0106) |
| Beyoncé | | | | | -0.424*** | | | |
| | | | | | (0.0179) | | | |
| Lebron James | | | | | | -0.594*** | | |
| | | | | | | (0.0131) | | |
| Lady Gaga | | | | | | | -0.302*** | |
| | | | | | | | (0.0121) | |
| Celebrity | | | | | | | | -0.391*** |
| | | | | | | | | (0.0107) |
| Year F.E. | Yes | Yes | Yes | Yes | Yes | Yes | Yes | Yes |
| Constant | 1.924*** | 1.767** | 1.172 | 1.588*** | 1.595*** | 1.619*** | 1.627*** | 1.654*** |
| | (0.534) | (0.540) | (1.125) | (0.413) | (0.413) | (0.413) | (0.413) | (0.414) |
| **Independent Follower (baseline)** | | | | | | | | |
| **Republican Follower** | | | | | | | | |
| Tweets | -25.61*** | -18.03*** | -14.99*** | -22.70*** | -21.87*** | -22.62*** | -22.66*** | -22.22*** |
| | (0.842) | (0.741) | (2.059) | (0.618) | (0.619) | (0.619) | (0.619) | (0.619) |
| Social Capital | -38.05*** | -22.61*** | -59.30** | -29.51*** | -27.58*** | -29.51*** | -27.82*** | -28.07*** |
| | (7.154) | (5.235) | (18.91) | (4.992) | (4.937) | (5.000) | (4.918) | (4.924) |
| Journalist | -1.838*** | -1.804*** | -1.506*** | -1.838*** | -1.851*** | -1.859*** | -1.858*** | -1.871*** |
| | (0.0911) | (0.0834) | (0.222) | (0.0716) | (0.0718) | (0.0717) | (0.0717) | (0.0718) |
| Name | -0.130*** | | | | | | | |
| | (0.0123) | | | | | | | |
| Image | | -0.0207 | | | | | | |
| | | (0.0148) | | | | | | |
| Description | | | -0.0888* | | | | | |
| | | | (0.0393) | | | | | |
| Female | | | | -0.0518*** | -0.00658 | -0.0895*** | -0.0237* | -0.0525*** |
| | | | | (0.0104) | (0.0104) | (0.0104) | (0.0104) | (0.0104) |
| Beyoncé | | | | | -0.895*** | | | |
| | | | | | (0.0180) | | | |
| Lebron James | | | | | | -0.406*** | | |
| | | | | | | (0.0123) | | |
| Lady Gaga | | | | | | | -0.577*** | |
| | | | | | | | (0.0118) | |
| Celebrity | | | | | | | | -0.545*** |
| | | | | | | | | (0.0103) |
| Year F.E. | Yes | Yes | Yes | Yes | Yes | Yes | Yes | Yes |
| Constant | 1.393* | 1.422* | 0.859 | 1.203** | 1.215** | 1.226** | 1.265** | 1.286** |
| | (0.582) | (0.575) | (1.226) | (0.449) | (0.449) | (0.449) | (0.450) | (0.450) |
| Observations | 387148 | 294987 | 30786 | 557777 | 557777 | 557777 | 557777 | 557777 |

Standard errors in parentheses
* $p < 0.05$, ** $p < 0.01$, *** $p < 0.001$

Table 9: Celebrity Effects: A Logistic Regresssion

| | Democrat | Democrat | Democrat | Democrat | Democrat | Clinton | Clinton | Clinton | Clinton |
|---|---|---|---|---|---|---|---|---|---|
| Social Capital | 36.23*** | 37.99*** | 37.96*** | 36.74*** | 36.81*** | 5.946 | 6.922 | 5.404 | 5.356 |
| | (5.194) | (5.273) | (5.272) | (5.209) | (5.211) | (4.887) | (4.866) | (4.833) | (4.705) |
| Journalist | 1.242*** | 1.230*** | 1.230*** | 1.245*** | 1.245*** | 1.057*** | 1.081*** | 1.084*** | 1.100*** |
| | (0.0620) | (0.0619) | (0.0619) | (0.0620) | (0.0619) | (0.103) | (0.103) | (0.103) | (0.103) |
| Female | 0.525*** | 0.531*** | 0.523*** | 0.534*** | 0.546*** | 0.0484*** | 0.123*** | 0.0534*** | 0.0753*** |
| | (0.00597) | (0.00598) | (0.00636) | (0.00595) | (0.00594) | (0.0105) | (0.0105) | (0.0105) | (0.0105) |
| Beyoncé | 0.489*** | | | | | 0.578*** | | | |
| | (0.0126) | | | | | (0.0226) | | | |
| Lebron James | | -0.183*** | -0.202*** | | | | 0.865*** | | |
| | | (0.00855) | (0.0102) | | | | (0.0185) | | |
| Female # James | | | 0.0654*** | | | | | | |
| | | | (0.0185) | | | | | | |
| Lady Gaga | | | | 0.283*** | | | | 1.012*** | |
| | | | | (0.00766) | | | | (0.0159) | |
| Celebrity | | | | | 0.163*** | | | | 0.783*** |
| | | | | | (0.00653) | | | | (0.0126) |
| Year F.E. | Yes | Yes | Yes | Yes | Yes | Yes | Yes | Yes | Yes |
| Constant | 0.578* | 0.595* | 0.597* | 0.562 | 0.565 | 0.606 | 0.593 | 0.513 | 0.509 |
| | (0.289) | (0.289) | (0.289) | (0.289) | (0.289) | (0.357) | (0.357) | (0.360) | (0.359) |
| Observations | 509810 | 509810 | 509810 | 509810 | 509810 | 227525 | 227525 | 227525 | 227525 |

Standard errors in parentheses

* $p < 0.05$, ** $p < 0.01$, *** $p < 0.001$

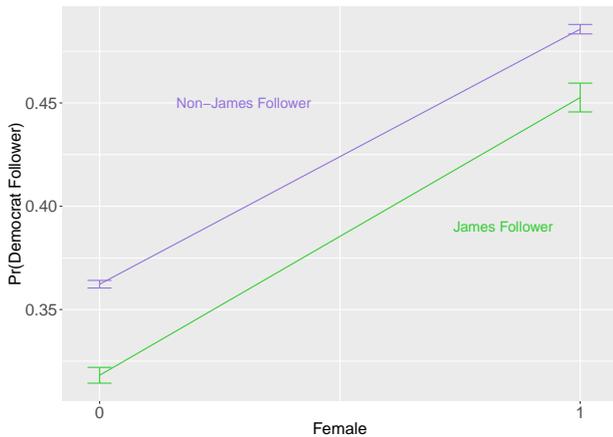

Figure 8: While Lebron James has endorsed Hillary Clinton, his followers are less likely to follow Democrats than individuals who do not follow him.

## Conclusion

This paper studies the paradoxical observation that while technically the marginal cost of following one extra presidential candidate is just a click a way, most individuals choose to follow only 1 or 2 candidates. Building from an exhaustive dataset that includes 15.5 million records, taking advantage of three information channels (name, image, description), and applying various regression models, we (1) explored the frequent patterns of the 2016 U.S. presidential campaign on Twitter, calculated the weighted presence for each candidate, and measured the extent to which individuals on Twitter are polarized, (2) studied how gender has had an effect on the number of candidates that one chooses to follow, (3) found that females are more likely to be following Democratic candidates exclusively and that followers of celebrities tend to be following candidates from both parties, (4) found that when considering Democrats and Republicans only, followers of Beyoncé and Lady Gaga are more likely to follow the Democratic candidates, that followers of Lebron James are more likely to follow the Republican candidates and that all of these tend to favor Hillary Clinton over Bernie Sanders.

## Acknowledgments

We gratefully acknowledge the generous financial support from the University and from our corporate sponsors.

## References

Alaimo, K. 2016. Where donald trump got his real power. *CNN*.

Barberá, P. 2015. Birds of the same feather tweet together. bayesian ideal point estimation using twitter data. *Political Analysis* 23(1).

Brians, C. L. 2005. Women for Women? Gender and Party Bias in Voting for Female Candidates. *American Politics Research*.

Burger, J. D.; Henderson, J.; Kim, G.; and Zarrella, G. 2011. Discriminating gender on twitter. *Proceedings of*

the 2011 Conference on Empirical Methods in Natural Language Processing.

Burt, R. S. 2000. Decay functions. *Social Networks* 22:1–28.

Campbell, J. E. 2016. *Polarized: Making Sense of a Divided America*. Princeton University Press.

Doherty, C. 2014. 7 things to know about polarization in america. *Pew Research Center*.

Dolan, K. 2008. Is There a "Gender Affinity Effect" in American Politics? Information, Affect, and Candidate Sex in U.S. House Elections. *Political Research Quarterly*.

Druckman, J. N.; Peterson, E.; and Slothuus, R. 2013. How elite partisan polarization affects public opinion formation. *American Political Science Review* 107(1):57–79.

Farfade, S. S.; Saberian, M.; and Li, L.-J. 2015. Multi-view face detection using deep convolutional neural networks. In *ICMR*.

Ginosar, S.; Rakelly, K.; Sachs, S.; Yin, B.; and Efros, A. A. 2015. A Century of Portraits: A Visual Historical Record of American High School Yearbooks. In *ICCV 2015 Extreme Imaging Workshop Proceedings*.

Greene, W. 2008. Functional forms for the negative binomial model for count data. *Economic Letters* (99):585–590.

Guo, Y.; Zhang, L.; Hu, Y.; He, X.; and Gao, J. 2016. Ms-celeb-1m: A dataset and benchmark for large-scale face recognition. *ECCV*.

Han, J.; Kamber, M.; and Pei, J. 2011. *Data Mining: Concepts and Techniques*. Morgan Kaufmann.

Hare, C., and Poole, K. T. 2014. The polarization of contemporary american politics. *Polity* 46(3):411–429.

He, K.; Zhang, X.; Ren, S.; and Sun, J. 2016. Deep residual learning for image recognition. *CVPR*.

Huang, G. B.; Ramesh, M.; Berg, T.; and Learned-Miller, E. 2007. Labeled faces in the wild: A database for studying face recognition in unconstrained environments. Technical report, University of Massachusetts.

Jia, S., and Cristianini, N. 2015. Learning to classify gender from four million images. *Pattern Recognition Letters*.

Jr., K. R., and Tesafaye, T. 2006. Morph: a longitudinal image database of normal adult age-progression. *7th International Conference on Automatic Face and Gesture Recognition (FGR06)*.

King, D. C., and Matland, R. E. 2003. Sex and the Grand Old Party: An Experimental Investigation of the Effect of Candidate Sex on Support for a Republican Candidate. *American Politics Research*.

Krizhevsky, A.; Sutskever, I.; and Hinton, G. E. 2012. Imagenet classification with deep convolutional neural networks. In *NIPS*.

LeCun, Y.; Bengio, Y.; and Hinton, G. 2015. Deep learning. *Nature*.

Levi, G., and Hassner, T. 2015. Age and Gender Classification using Deep Convolutional Neural Networks. In *Proceedings of the IEEE Conference on Computer Vision and Pattern Recognition*, 34–42.

Lockhart, K. 2016. Watch: Why social media is donald trump's most powerful weapon. *The Telegraph*.

Maddala, G. S. 1983. *Limited-Dependent and Qualitative Variables in Econometrics*. Cambridge University Press.

McCarty, N.; Poole, K. T.; and Rosenthal, H. 2009. Does gerrymandering cause polarization? *American Journal of Political Science* 53(3):666.

McPherson, M.; Smith-Lovin, L.; and Cook, J. M. 2001. Birds of a feather: Homophily in social networks. *Annual Review of Sociology*.

Miller, C. C. 2016. Why women did not unite to vote against donald trump. *The New York Times*.

Mislove, A.; Lehmann, S.; Ahn, Y.-Y.; Onnela, J.-P.; and Rosenquist, J. N. 2011. Understanding the demographics of twitter users. *Proceedings of the Fifth International AAAI Conference on Weblogs and Social Media*.

Nilizadeh, S.; Groggel, A.; Lista, P.; Das, S.; Ahn, Y.-Y.; ; Kapadia, A.; and Rojas, F. 2016. Twitter's glass ceiling: The effect of perceived gender on online visibility. *Proceedings of the Tenth International AAAI Conference on Web and Social Media*.

Ottoni, R.; Pesce, J. P.; Casas, D. L.; Jr., G. F.; Jr., W. M.; Kumaraguru, P.; and Almeida, V. 2013. Ladies first: Analyzing gender roles and behaviors in pinterest. *Proceedings of the Seventh International AAAI Conference on Weblogs and Social Media*.

Phillips, P. J.; Wechslerb, H.; Huangb, J.; and Raussa, P. J. 1998. The feret database and evaluation procedure for face-recognition algorithms. *Image and Vision Computing* 295–306.

Sanders, B. 2016. *Our Revolution: A Future to Believe In*. Thomas Dunne Books.

Simonyan, K., and Zisserman, A. 2015. Very Deep Convolutional Networks for Large-Scale Image Recognition. In *International Conference on Learning Representations 2015*.

Srivastava, R. K.; Greff, K.; and Schmidhuber, J. 2015. Highway network. *arXiv:1505.00387v2*.

Stahl, L. 2016. President-elect trump speaks to a divided country on 60 minutes. *CBS*.

Tumasjan, A.; Sprenger, T. O.; Sandner, P. G.; and Welpe, I. M. 2010. Predicting Elections with Twitter: What 140 Characters Reveal about Political Sentiment. In *Proceedings of the Fourth International AAAI Conference on Weblogs and Social Media*.

Wang, Y.; Feng, Y.; Zhang, X.; and Luo, J. 2016. Voting with Feet: Who are Leaving Hillary Clinton and Donald Trump? In *Proceedings of the IEEE Symposium on Multimedia*.

Wang, Y.; Li, Y.; and Luo, J. 2016. Deciphering the 2016 U.S. Presidential Campaign in the Twitter Sphere: A Comparison of the Trumpists and Clintonists. In *Tenth International AAAI Conference on Web and Social Media*.

Zamal, F. A.; Liu, W.; and Ruths, D. 2012. Homophily and latent attribute inference: Inferring latent attributes of twitter users from neighbors. *Proceedings of the Sixth International AAAI Conference on Weblogs and Social Media*.